\documentclass[12pt]{iopart}
\usepackage{graphicx}
\usepackage{cite}

\eqnobysec

\begin{document}

\title[]
{Statistics of close-packed dimers on fractal lattices}

\author{D Mar\v{c}eti\'{c}$^1$, S Elezovi\'c-Had\v zi\'c$^2$ and I \v{Z}ivi\'{c}$^3$ }

\address{$^1$ Faculty of Natural Sciences and Mathematics, University of Banja Luka, M.~Stojanovi\'{c}a 2, Bosnia and Herzegovina}
\address{$^2$ Faculty of Physics,
University of Belgrade, P.O.Box 44, 11001~Belgrade, Serbia}
\address{$^3$ Faculty of  Science, University of Kragujevac, Radoja Domanovi\'{c}a 12, Kragujevac, Serbia}
\eads{\mailto {dusanka.marcetic-lekic@pmf.unibl.org}, \mailto{suki@ff.bg.ac.rs},  \mailto{ivanz@kg.ac.rs}}

\begin{abstract}
 We study the model of  close-packed dimers  on planar lattices belonging to the family of modified rectangular (MR) fractals, whose members are enumerated by an integer $p\geq 2$, as well as  on the non-planar 4-simplex fractal lattice. By applying an exact recurrence enumeration method, we determine the asymptotic forms for numbers of dimer coverings, and numerically calculate entropies per dimer in the thermodynamic limit, for a sequence of MR lattices with $2\leq p\leq8$ and for 4-simplex fractal. We find that the entropy per  dimer  on MR  fractals   is increasing function of the scaling  parameter $p$, and for every considered $p$ it  is smaller than the  entropy per dimer of the same model  on $4$-simplex lattice.  Obtained results are discussed and compared with the results obtained previously on some translationally invariant  and fractal lattices.
\end{abstract}

\vskip 5mm


\maketitle

\section{Introduction}
\label{prva}

A close-packed dimer model is a classical  example of a lattice statistical mechanical model that can be solved exactly on regular planar lattices. It has emerged as a simplified  version of the  so called  monomer-dimer model, that was  introduced in 1937 by Fowler and Rushbroke \cite{Fowler}  in their   study of the  liquid mixtures. Exact solution of the  close-packed dimer model on  the square lattice was given  in the $60$-ies    by Kasteleyn \cite{Kasteleyn1,Kasteleyn2}, and independently by Temperly and Fisher \cite{Temperley,Temperley1} and Fisher    \cite{Fisher}. More recent solution on  arbitrary planar bipartite graphs has been given by Kenyon et al \cite{Kenyon}.  Although introduced as a  simple model  for adsorption of diatomic molecules on crystal surfaces,   connections with  other models in physics and chemistry have been established since then. It has been shown that the close-packed dimer model is  equivalent to  the two dimensional Ising model \cite{Fisher2}, and correspondence with   many  quantum theoretical field theory models  has been recognized  \cite{Nienhuis, Kondev, Iqbal, Jacobsen, Jacobsen1}.  In graph theory the  model is also referred to  as perfect matchings and  is closely related with other combinatorial objects such as    domino tillings \cite{ Cohn} and spaning trees \cite{ Kenyon2,Temperley2}.

Besides the square lattice, the close-packed dimer  model has also been studied on other translationally invariant lattices \cite{Elser,Fendley, Wu},  on  some graphs without translational symmetry, such as  self-similar graphs \cite{Harris, Dangeli}, and particularly  on   the  Sierpinski gasket lattice and its generalizations  \cite{Chang}.

In this paper we analyze close-packed dimer model on  the subset of  planar fractal lattices that belong to the family of the  modified rectangular  (MR) lattice introduced by Dhar \cite{Dhar}. We also  consider the model  on 4-simplex lattice which is  a fractal lattice  embedded in three dimensional space  and whose graph is consequently non-planar. Fractal lattices are   constructed iteratively, which  makes them   suitable for exact recursive treatment if the ramification number is low.   Recursive enumerative  method  on these lattices can  provide exact solutions for the close-packed dimer model, in addition to solutions obtained   on translationally invariant planar lattices,  which proved to be analytically tractable in that case. However,  there are no exact solutions of the  close-packed dimer model on   three dimensional translationally invariant lattices, and result   obtained  on  $4$-simplex fractal lattice in this paper, together with  the results on other fractal lattices  embedded in three dimensional space \cite{Dhar2, Chang},   can  give valuable insights into the problem essence.

 The paper is organized as follows. In section~\ref{druga} we shortly  describe the  closed-packed dimer model and fractal lattices relevant for this paper. In section~\ref{treca} we develop recursive method for the enumeration of all dimer configurations on MR family for scaling parameter $2\leq p\leq8$ and analyze the model to obtain the entropy. In the same section, we also  apply the method on  4-simplex lattice.  Discussion of the results and comparison with other  lattices are given in section~\ref{cetvrta}.

\section{The model and lattices}
\label{druga}

We assume that dimer is a diatomic molecule, i.e. two  monomer units  bonded chemically. On a lattice it  covers  two adjacent lattice points. In close-packed dimer model each lattice site is occupied exactly once by a monomer, and   each monomer is connected    by a lattice bond  with  an  adjacent monomer  into a dimer.  Neglecting  any  other interactions besides hard-core repulsion,   partition function of this model is simply  the total number of dimer configurations, whose logarithm determines the entropy.
\begin{figure}
\begin{center}
\includegraphics[scale=0.95]{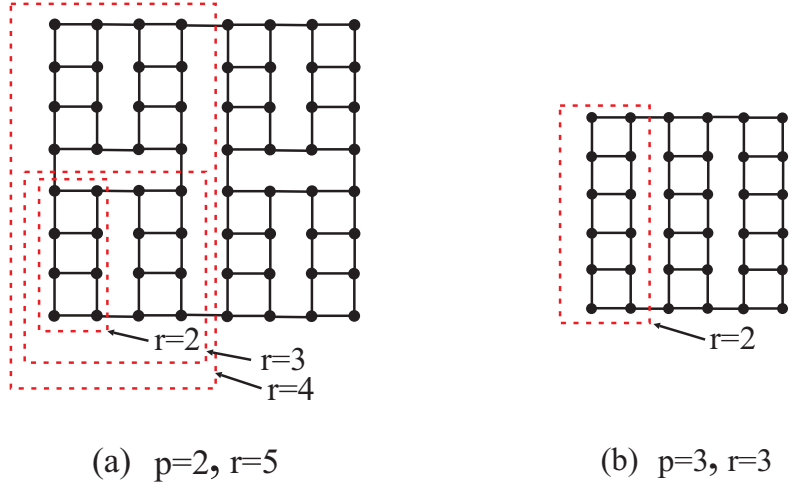}
\end{center}
\caption{(a) Generator of order $r=5$ for  MR  lattice  with $p=2$. Red  dashed rectangles highlight  generators of order $r=2$, $r=3$, and $r=4$ obtained in  the subsequent stages of construction.    (b)   Generator of order $r=3$  for  $p=3$  MR fractal, with  smaller  sub-generator (of order $r=2$) highlighted by red dashed rectangle. In both cases generator of order $r=1$ (initiator) is a unit square.}
 \label{fig1}
\end{figure}
 Lattices under consideration are   MR family of fractals   and $4$-simplex lattice.
  Fractals from MR family are labeled  with  the  scaling parameter  $p$ (an integer,  $2\leq p\leq \infty$).  In iterative constructive procedure, structure obtained in the construction step $r$ is called  $r$-th order generator and   denoted by $G_r$. For each particular $p$, at the first step  of  construction  ($r=1$) one has a  graph consisting of  four points forming a unit square. Then, $p$ unit squares are joined into the rectangle to obtain  the generator of the second  order. In the next step, $p$ rectangles are joined into the  square, and the process should be repeated  infinitely many times to obtain fractal lattice.   In figure~\ref{fig1}$(a)$ generator of order $r=5$ for  $p=2$ MR lattice  is shown, while in figure~\ref{fig1}$(b)$ generator of order $r=3$  for $p=3$ member of MR family is shown.  Generator of order $r$ for each fractal   contains $N_{r}=4p^{\,r-1}$ lattice sites (it is also the number of monomers - twice  the number of dimers, because of close-packing) and  $N_{br}=\frac{3}{2}N_r-2=\frac{6}{p}p^{\,r}-2$  lattice bonds (edges).   Fractal dimension is $d_f=\frac{\ln p^2}{\ln p}=2$ for each fractal  from the family.
  \begin{figure}
\begin{center}
\includegraphics[scale=0.7]{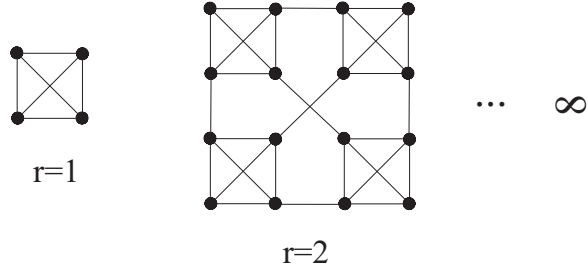}
\end{center}
\caption{First two steps of the iterative construction of $4$-simplex lattice. Fractal lattice   is obtained in the limit $r\rightarrow\infty$.}
 \label{fig2}
\end{figure}

 Initiator  of   $4$-simplex lattice graph   is a complete graph of four points. To obtain second order generator, four initiators  are joined into two times   larger structure, as shown in figure~\ref{fig2}, and the process should be repeated ad infinitum to obtain  graph of $4$-simplex lattice. The number of lattice points in the $r$-th order generator is $N_r=4^r$,  whereas the number of lattice bonds is $N_{br}=\frac{4}{2}N_r-2=2\cdot4^r-2$. Fractal dimension of the lattice is $d_f=\frac{\ln 4}{\ln 2}=2$.

 It should be emphasized that for all considered lattices the number of lattice points in  generators of any order is even, which is  a necessary  condition for a lattice to have  close-packed dimer covering.

\section{Recursive enumerative method  for dimer coverings on fractal lattices   \label{treca}}

In this section we will develop  the method for recursive enumeration  of dimer configurations on aforementioned  fractal lattices.   Firstly, we  establish  the exact set of recurrent equations on  MR lattices with $2\leq p\leq 8$  and analyze equations in order to  determine the asymptotic form for the numbers of  dimer coverings. Also,  we  numerically  find  the corresponding entropies per dimer in the thermodynamic limit.  Although   for some lattice models  it was possible to find exact set of   recurrence equations on the whole  MR family \cite{dusa}, in this case  we were not able to do so   for the reason that would be explained in the appendix~A. Secondly,  we  apply the same method  on $4$-simplex lattice and  determine the entropy.
\subsection{Dimer coverings on MR  fractals   \label{dimeriMR}}

\begin{figure}
\begin{center}
\includegraphics[scale=0.7]{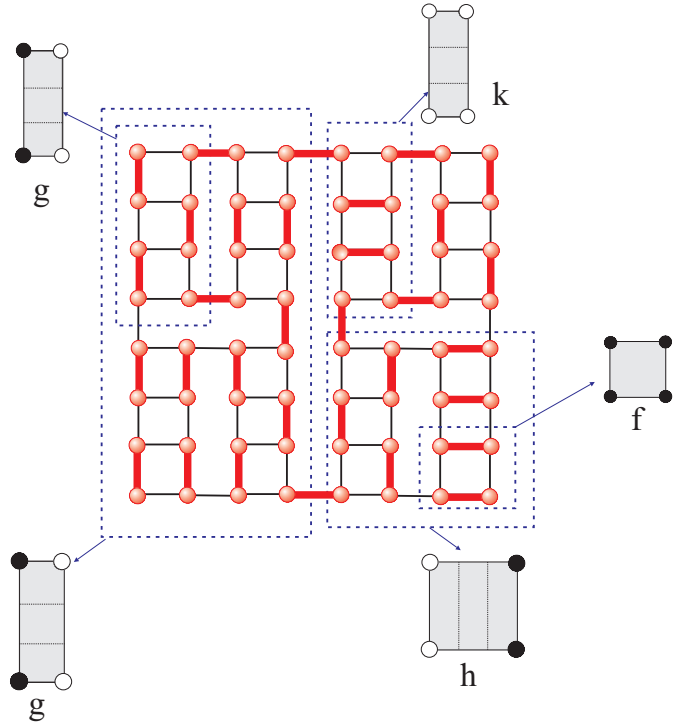}
\end{center}
\caption{One  close-packed dimer configuration on  generator $G_5$ of  $p=2$ MR lattice. Dimer configurations on some sub-generators (rectangles or squares) are enclosed  into blue  dashed rectangles (squares), outlined   on the sides and labeled as $f$, $g$, $h$ or $k$,  depending  on the  type of the configuration.}
 \label{fig3}
\end{figure}

 One close-packed dimer configuration on the $5$-th order generator of $p=2$ MR lattice is shown in figure~\ref{fig3}. In order to develop recurrence equation for the number of dimer coverings on MR lattice,  we  focus on  the corner monomers of smaller  generators, that is $G_4$, $G_3$, $G_2$, and $G_1$,  as sub-generators of $G_5$ depicted  in figure~\ref{fig3}.  One can notice  that the corner monomers  of generators of any order  form  dimers either by the monomers on the same generator or the neighboring ones. We designate the corner monomers as  black if their partner is on the same generator, and white if it is on the neighboring one. All generators  have four   corner monomers and, due to the  parity, the only possible combinations are to have all four black, two black and two white, and all four white. Configurations with two black  and two white along the diagonals are not possible on MR lattices of any $p$.
According to the pairing of the  corner monomers, for MR lattices of arbitrary $p$,   we introduce four types of configurations on generators of any order $r$, namely $f$, $g$, $h$ and $k$ with the following  meaning:
\begin{itemize}
                         \item $f$ - denotes each  dimer configuration in  which all four corner monomers are black. They  form dimers with the 'internal' monomers, i.e.  monomers on the same generator,
                         \item $g$ - denotes each dimer configuration in which two corner monomers are black and belong to the \textbf{different} \textbf{ sub-generators} of order $(r-1)$. These two black  corner monomers form dimers with the monomers on the same $G_r$, while the other two white  corner monomers  form dimers with the monomers on the two neighboring $G_r$,
                         \item $h$ - denotes each dimer configuration in which two corner monomers are black and belong to the \textbf{same} \textbf{ sub-generator} of order $(r-1)$. As in type $g$, these two black  corner monomers form dimers with the monomers on the same $G_r$, while the other two white  corner monomers  form dimers with the monomers on the two neighboring $G_r$,
                         \item $k$ - denotes each  dimer configuration in which all four corner monomers are white. They  form dimers with the 'external' monomers,  i.e. monomers  on the neighboring  generators.
                       \end{itemize}
 \begin{figure}
\begin{center}
\includegraphics[scale=0.75]{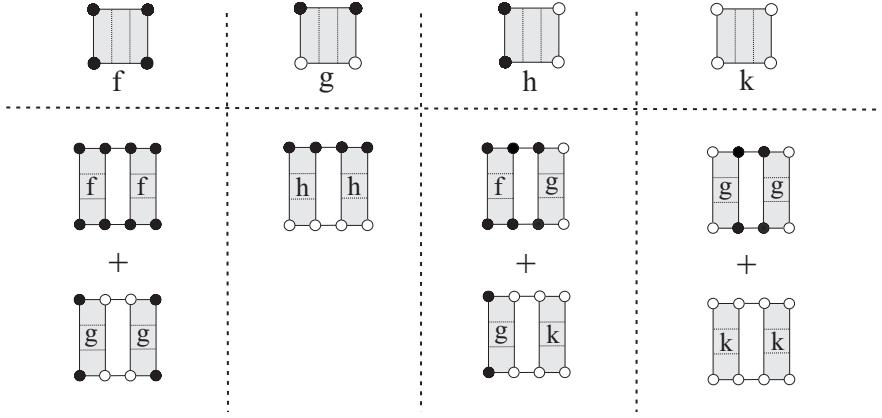}
\end{center}
\caption{ All types of  dimer  configurations on  arbitrary  generator $G_{r+1}$ of $p=2$ MR fractal, denoted as $f$, $g$, $h$ and $k$, and their composing configurations on generators $G_{r}$, from which recurrence equations (\ref{re1}) stem. }
 \label{fig4}
\end{figure}
In figure~\ref{fig3}, all four types of configurations are enframed     and schematically represented on the sides. In this schematic representation only corner monomers   are shown with the  internal structure of generators omitted, except that the two consecutive sub-generators are indicated by the dashed lines in order to easily distinguish between $g$ and $h$ configurations.

\begin{figure}
\begin{center}
\includegraphics[scale=0.95]{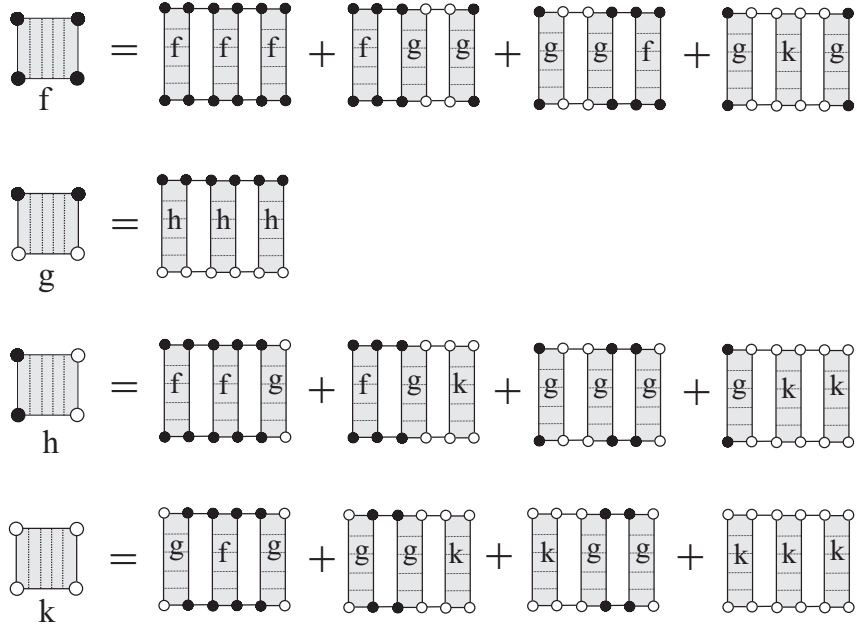}
\end{center}
\caption{Configurations $f$, $g$, $h$, and $k$ on  generator $G_{r+1}$ of $p=3$  member of MR fractals  and their composing parts on generators $G_r$. }
 \label{fig5}
\end{figure}

The  numbers of   configurations of each  type  on the  $r$-th order generator  are designated as $f_r$, $g_r$, $h_r$ and $k_r$. The total number of dimer configurations on $G_r$ is given by $f_r$, and  can be determined through the system of recurrence equations that involve all four configurations.  The closed system of the recurrence equations for $p=2$ is given as
\begin{eqnarray}
 f_{r+1}&=&f_r^2+g_r^2\, ,   \nonumber\\
 g_{r+1}&=&h_r^2\, , \nonumber \\
 h_{r+1}&=&f_r g_r+ g_r k_r\, ,  \nonumber\\
  k_{r+1}&=&g_r^2+k_r^2\, , \label{re1}
\end{eqnarray}
and can be inferred on the basis of  figure~\ref{fig4}, where we illustrate how each  configuration on the $(r+1)$-th order generator can be composed from the configurations on the two constituent $r$-th order generators.
Similarly, with the help of figure~\ref{fig5}, one can formulate recurrence equations for $p=3$ as
\begin{eqnarray}
 f_{r+1}&=&f_r^3+2f_r g_r^2+g_r^2k_r\, ,   \nonumber\\
 g_{r+1}&=&h_r^3\, , \nonumber \\
 h_{r+1}&=&f_r^2 g_r+f_r g_r k_r+g_r^3+g_r k_r^2\, ,  \nonumber\\
  k_{r+1}&=&f_r g_r^2+2g_r^2k_r+k_r^3\, .  \label{re2}
\end{eqnarray}
Recurrence equations on fractals with $4\leq p\leq8$ are given in appendix~A.  The initial conditions of these equations  are  the numbers of  configurations   on the unit square, and, for each $p$ they   are given by: $f_1=2$ (both two possible), $g_1=1$ (only one of the two possible), $h_1=1$ (only one of the  two possible) and $k_1=1$ (the only possible). By computer iteration of  recursion relations~(\ref{re1}), and similarly  ~(\ref{re2}),  starting with the  initial conditions,  it is possible to obtain exact numbers of dimer configurations on generators of arbitrary order.
  These numbers  grow very fast with the order  $r$ of generator, and  for illustration,  in table~\ref{tab1} we give the numbers of close-packed dimer configurations on generators of order from  $r=1$  to $r=5$ for MR fractals with $p=2$ and $p=3$.
\begin{table}
\caption{The numbers of closed-packed dimers on the first five generators for MR fractals labeled by  $p=2$ and $p=3$.}
\centering
\label{tab1}
\vspace{2mm}
\begin{tabular}{|c|ccccc|}\hline
 &$r=1$& $r=2$ & $r=3$ & $r=4$&$r=5$ \\ \hline\hline
$p=2$& 2 & 5& 26&757&575450\\
$p=3$&2 & 13& 2228&12266667328&1845787045627790291334622871552\\ \hline
 \end{tabular}
\end{table}
Since the systems of difference equations given by~(\ref{re1})   and  (\ref{re2}) are not   solvable  exactly, it is not possible to   find exact  expressions for $f_r$ as  functions of $r$. Therefore,  we  analyze relations numerically  and find asymptotic  solutions. To make numerical analysis more tractable, we introduce new, rescaled variables defined as $x_r=g_r/f_r$, $y_r= h_r/f_r$ and $z_r= k_r/f_r$, whose recurrence equations  can be obtained from  their definitions and the equations~(\ref{re1}) and   (\ref{re2}).  New equations are
\begin{eqnarray}
 f_{r+1}&=&f_r^2\left(1+x_r^2\right)\, ,   \nonumber\\
 x_{r+1}&=&\frac{y_r^2}{1+x_r^2}\, , \nonumber \\
 y_{r+1}&=&\frac{x_r\left(1+z_r\right)}{1+x_r^2}\, ,  \nonumber\\
  y_{r+1}&=&\frac{x_r^2+z_r^2}{1+x_r^2}\, , \label{re1x}
\end{eqnarray}
for $p=2$,   and
\begin{eqnarray}
 f_{r+1}&=&f_r^3\left(1+2x_r^2+x_r^2z_r\right)\, ,   \nonumber\\
 x_{r+1}&=&\frac{y_r^3}{1+2x_r^2+x_r^2z_r}\, , \nonumber \\
 y_{r+1}&=&\frac{x_r\left(1+z_r+x_r^2+z_r^2\right)}{1+2x_r^2+x_r^2z_r}\, ,  \nonumber\\
  y_{r+1}&=&\frac{x_r^2+2x_r^2z_r+z_r^3}{1+2x_r^2+x_r^2z_r}\, , \label{re2x}
\end{eqnarray}
for $p=3$.
Initial values of new variables are $x_1=y_1=z_1=\frac{1}{2}$.

For arbitrary $p$,  the recurrence equation for the number $f_r$ of close-packed dimers as a function of rescaled variables can be written as
\begin{equation}\label{fpx}
f_{r+1}=f_r^p\left(1+\sum_{i}a_ix_i^{\alpha_i}z_i^{\beta_i}\right)\, ,
\end{equation}
 where  the coefficients $a_i$  and the exponents $\alpha_i$ and $\beta_i$  depend on  $p$. First equation in systems~(\ref{re1x}) and (\ref{re2x}) are indeed  of  the given form.  Iterating  sequences $x$, $y$ and $z$,  we find that for all considered $2\leq p\leq8$,  elements  $x_r$, $y_r$ and $z_r$  tend to zero very quickly with each iteration step  $r$, implying that the equation~(\ref{fpx}) for $r\gg1$ has the form
\begin{equation}\label{fpas}
f_{r+1}\sim f_r^p\, .
\end{equation}
This further implies that $f_r$ asymptotically grows exponentially with $p^{\,r}$ i.e. $f_r\sim[const]^{p^{\,r}}$. Since the number of  monomers in $G_r$ is given by the $N_r=\frac{4}{p}p^{\,r}$  it follows that  $f_r$  exponentially grows with the number of monomers
\begin{equation}\label{fpas1}
f_{r}\sim\omega^{N_r}\, ,
\end{equation}
\begin{table}
\caption{Entropies  per dimer  $s_d$  of close-packed dimer model on  MR fractals with $2\leq p\leq 8$. The last digits are rounded off.}
\centering
\label{tab2}
\vspace{2mm}
\begin{tabular}{c|ccccccc}
\hline\hline
${ p}$ &2 & 3 & 4 &5 \\ \hline
$s_d$ & 0.414750739 & 0.430188671 & 0.441389262 & 0.449006803\\ \hline\hline
${ p}$ &6 & 7 & 8 & - \\ \hline
$s_d$& 0.454290896 & 0.458114819 & 0.460997823 &-\\  \hline\hline
 \end{tabular}
\end{table}
 for $r\gg1$. Growth constant $\omega$ is defined through the relation $\ln\omega=\lim_{N_r\to\infty}\frac{\ln f_r}{N_r}$. To determine growth constant, we take  logarithm of the equation (\ref{fpx}) and  divide obtained equation with $N_{r+1}=4p^{\,r}$,  after which we obtain
\begin{equation}\label{nizs}
\frac{\ln f_{r+1}}{N_{r+1}}=\frac{\ln f_r}{N_r}+\frac{1}{4p^{\,r}}\ln\left(1+\sum_{i}a_ix_i^{\alpha_i}z_i^{\beta_i}\right)\, .
\end{equation}
\begin{figure}
\begin{center}
\includegraphics[scale=0.4]{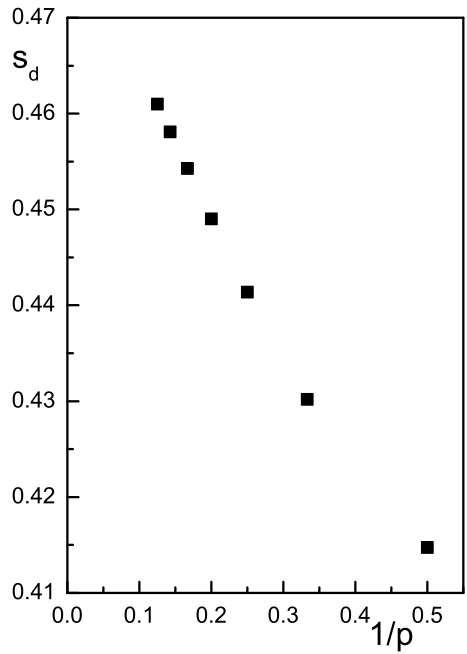}
\end{center}
\caption{Entropies per dimer of  the close-packed dimer model on MR lattices with $2\leq p\leq8$ as functions of $1/p$. }
 \label{fig7}
\end{figure}
 This equation recursively  defines  sequence of numbers with the elements  given by  $s_r=\frac{\ln f_r}{N_r}$, so that $s_m=\lim_{r\to\infty} s_r=\ln\omega$ holds. For each $p$, the sequence converges very fast, so that for example for $p=3$, after only nine iterations more than twenty significant figures can be achieved, and for higher $p$ convergence is even faster. By numerical iteration, value of $s_m$  is determined  for  each $2\leq p\leq 8$. Limiting values $s_m$ are actually  the entropies per monomer in  the thermodynamic limit, as can be seen from the definition of the entropy through the  number of dimer configurations $S=k_B\ln f_r$ and  equation~(\ref{fpas1}).   Setting $k_B=1$,  it follows that $\lim_{N_r\to\infty} S/N_r=\ln \omega=s_m$. The number of dimers is just half of the number of monomers $N_r$, so that the entropy per dimer $s_d$ is twice the entropy per monomer. In table~\ref{tab2} we present entropies per dimer in the thermodynamic limit, calculated  numerically from the numbers of dimer configurations on MR   fractals  with $2\leq p\leq 8$. Entropies  $s_d$ as functions of $1/p$
are  also   presented graphically in figure~\ref{fig7}.

\subsection{Dimer coverings on $4$-simplex lattice   \label{dimeri4sim}}

\begin{figure}
\begin{center}
\includegraphics[scale=0.7]{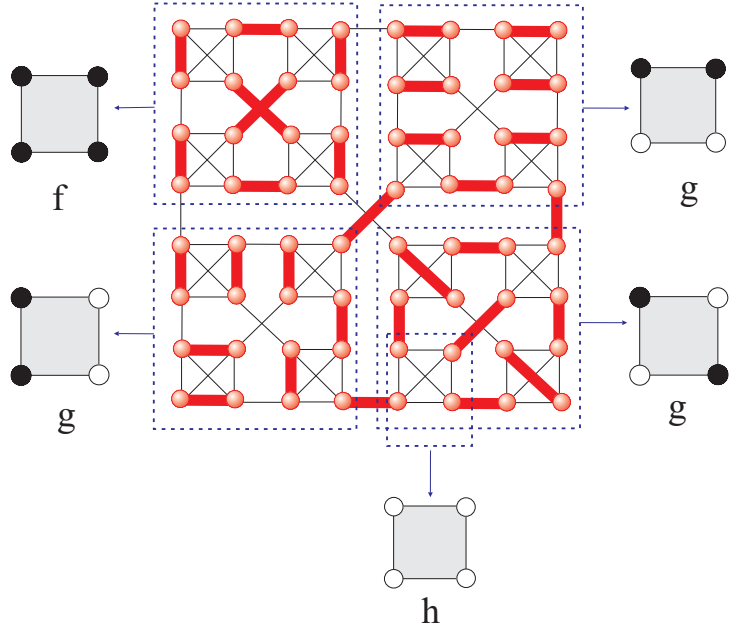}
\end{center}
\caption{Close-packed dimer configuration on the generator $G_3$ of $4$-simplex lattice. Possible types of  configurations on sub-generators are enframed and  schematically shown on the sides. }
 \label{fig8}
\end{figure}
\begin{figure}
\begin{center}
\includegraphics[scale=0.95]{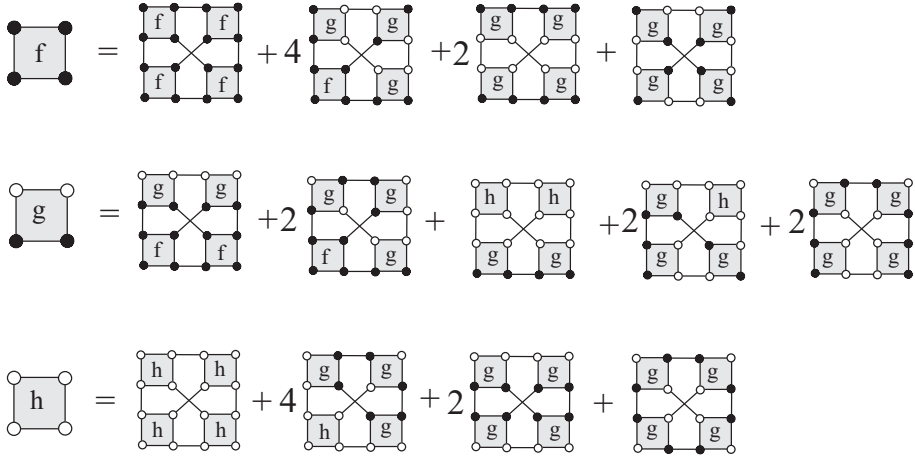}
\end{center}
\caption{Configurations $f$, $g$ and $h$ on  generator $G_{r+1}$ of $4$-simplex lattice and their composing parts on generators $G_r$,  which  determine corresponding  terms in equations~(\ref{resim}). Multiplication by the   factors $4$ or $2$  are due to the symmetrically related configurations. }
 \label{fig9}
\end{figure}

 Recursive enumeration of dimer configurations on $4$-simplex lattice can be done   in a  similar manner  as on MR lattices. In figure~\ref{fig8} one close-packed dimer configuration on the  third order generator of $4$-simplex lattice is presented.
Again, corner monomers can be all   four black - corresponding to $f$ configuration,  two black and two white - corresponding  to  $g$ configuration and all four white - corresponding to  $h$ configuration. All four corner vertices of  $4$-simplex lattice  are permutationally equivalent, so it is irrelevant which two monomers  are black (white).   Four  constitutive sub-generators of $G_3$ in figure~\ref{fig8} are enframed,  with the  configurations  classified  according to the type they belong to, and  schematically represented on the sides. Also, one configuration of type $h$ on $G_1$ is  enframed, and schematically represented on the side. Recurrence equations for all three configurations, as illustrated in figure~\ref{fig9},  are given as
\begin{eqnarray}
 f_{r+1}&=&f_r^4+4f_r g_r^3+3g_r^4\, ,   \nonumber\\
 g_{r+1}&=&f_r^2g_r^2+2f_r g_r^3+g_r^2h_r^2+2g_r^3h_r+2g_r^4\, , \nonumber \\
 h_{r+1}&=&h_r^4 + 4g_r^3h_r+3g_r^4\,  , \label{resim}
\end{eqnarray}
with the initial conditions $f_1=3$ (all three possible), $g_1=1$ (one of six possible) and $h_1=1$ (the only possible).
Analysis of recurrence equations~(\ref{resim}) proceeds in a similar way as on MR lattices, and we find that the number of all close-packed dimer configurations grows with the number of monomers as
\begin{equation}\label{fsim}
f_{r}\sim\omega^{N_r}\, ,
\end{equation}
with the entropy per dimer $s_d=2\ln\omega=0.571832556 $.
\section{Discussion and conclusions} \label{cetvrta}
We have studied  the close-packed dimer model on two types of fractal lattices, namely, lattices  from  MR family  with the scaling parameter  $2\leq p \leq 8$   embedded in two dimensional space, and    $4$-simplex lattice  embedded in three dimensional space.  The asymptotic forms for the number of dimer configurations on these lattices have been determined.   It is found that, on all considered fractals,   the asymptotic form for number of dimer configurations is a pure exponential function of the number of monomers.   In addition,  microcanonical   entropies per dimer in the thermodynamic limit  are  determined numerically. On MR lattices,  entropy is an  increasing function of fractal parameter $p$, as can be seen  in table~\ref{tab2}.  This deserves some insight into the  geometry of the lattices. All lattices from this family are constructed iteratively through succession  of generators, and  all have the same fractal dimension. The ramification number  of each lattice is  two, and for all latices  coordination number of each vertex is three  (vertex degree),  except for the corner vertices of the  largest generator. But, the number of bonds and their distribution are different,  and   lattices with higher value of $p$ at each construction stage have  larger number of bonds (caused by the larger number of vertices), so the  configurational space is larger.  In figure~\ref{fig7} entropies are shown as functions of inverse scaling parameter $p$, and an   obvious question arises: is this sequence convergent when $p\rightarrow\infty$, and if so, what is the limit? We expect that there is a  finite limiting value, and suppose that it is smaller than the value of entropy $s_{sq}=0.583121808$ obtained on the square lattice in \cite{Kasteleyn1}. This conclusion is justified by the fact  that all MR lattices  resemble to  the square lattice since they  can be obtained from it by deleting some bonds. Consequently, MR lattices have   smaller number of bonds  than   the square lattice of equal size.

 On the other hand,   entropy obtained on $4$-simplex lattice  is $s_{sim}=0.571832556$, and compared with the values  in table~\ref{tab2}, we see that it is larger than any value on MR fractals considered. Coordination number of  $4$-simplex lattice is four and thus greater than for MR lattices, but for MR fractals with $p\geq5$ at $r$-th construction stage there are more bonds than  on  $4$-simplex lattice at the same stage. However, MR lattices are  highly anisotropic and their connectedness does not  allow some configurations (in subsection~\ref{dimeriMR}   configurations with two black and two white along the square (rectangle) diagonals  were forbidden), resulting in smaller entropy.
Value $s_{sim}$ obtained here, should also be compared  with the value of entropy of the same model studied on  the  Sierpinski gasket (SG) embedded in three dimensional space\cite{Chang}, which is $s_{3dSG}=0.857927798$, thus much larger than $s_{sim}$.  These two lattices both have tetrahedral structure, but vertices of  the neighboring tetrahedra are glued on $3d$ SG, so that the  coordination  number  of this lattice is six.
  Furthermore, entropy per dimer  is larger on  square lattice than on   $4$-simplex lattice, although they  both have the same coordination number.

  To conclude, we can confirm that besides the coordination number, which  is the most  relevant lattice parameter  that determines the number of close-packed dimer configurations,  other lattice  parameters and geometric constraints  are important too. In favor of latter is the finding that on translationally invariant lattices  the boundary effects play an important role. For example, on square lattice the entropy is the same for both  open  and  periodic  boundary conditions \cite{Kasteleyn1, Fisher}, whereas on hexagonal lattice it  strongly depends on the boundary \cite{Grensing, Elser}. Extensions of the studies to account for  cylindrical boundary conditions have also been done \cite{Yan, Li}. Similar problem has been encountered on other 'close-packed' models, namely Hamiltonian walk problem, where entropic exponent $\gamma$ depended on the boundary conditions\cite{Duplantier}. With respect to all these unresolved questions, we can say that  additional studies should be conducted in order to specify  all relevant parameters  on fractal lattices as well as on  translationally invariant ones,  taking into account all metric properties of lattices.

 Finally, we would like to  mention that the  dimer model  considered here  could be supplemented with the interaction weights and  studied on  fractal lattices. Interacting dimer model is  extensive  and physically more interesting, but also more difficult  to approach.

\ack{This paper has been done as a part of the
work within the project No. 171015, funded by  the Ministry of Education, Science and Technological Development of the Republic of Serbia.
}

\appendix

\section{}

In this appendix we  give recurrence equations for the determination of the number of dimer coverings on MR  fractals with $4\leq p\leq8$. We  explicitly state  the recurrence equations only for the variables $f$ and $h$, since the recurrence relation for the configuration $g$ is simply $g_{r+1}=h_r^p$ for each  $p$, and for $k$ it could be obtained from the recurrence equation for $f$ with the substitution $k\leftrightarrow f$ on both sides of equation.

$p=4:$
\begin{eqnarray}
\fl f_{r+1}&=&f_r^4+3f_r^2g_r^2+2f_r g_r^2k_r+g_r^4+g_r^2k_r^2\, ,   \nonumber\\
\fl h_{r+1}&=&f_r^3g_r+f_r^2g_r k_r+2f_r g_r^3 +f_r g_r k_r^2+2g_r^3k_r+g_r k_r^3\, . \label{ap1}
\end{eqnarray}\\

$p=5:$
\begin{eqnarray}
\fl f_{r+1}&=&f_r^5+4f_r^3g_r^2 +3f_r^2g_r^2k_r+3f_r g_r^4 +2f_r g_r^2k_r^2+2g_r^4k_r+g_r^2k_r^3\, ,   \nonumber\\
\fl h_{r+1}&=&f_r^4g_r+f_r^3g_r k_r+3f_r^2 g_r^3 +f_r^2 g_r k_r^2+4f_r g_r^3k_r+g_r^5+f_r g_r k_r^3+3g_r^3k_r^2+g_r k_r^4\, . \label{ap2}
\end{eqnarray}\\

$p=6:$
\begin{eqnarray}
\fl f_{r+1}&=&f_r^6+5f_r^4g_r^2 +4f_r^3g_r^2k_r+6f_r^2g_r^4 +3f_r^2g_r^2k_r^2+6f_r g_r^4k_r+g_r^6+2f_r g_r^2k_r^3+3g_r^4k_r^2\nonumber\\
\fl &+&g_r^2k_r^4\, ,   \nonumber\\
\fl h_{r+1}&=&f_r^5g_r+f_r^4g_r k_r+4f_r^3 g_r^3 +f_r^3 g_r k_r^2+6f_r^2 g_r^3k_r+3f_r g_r^5+f_r^2 g_r k_r^3+6f_r g_r^3k_r^2 \nonumber\\
\fl&+&3g_r^5k_r+f_r g_r k_r^4+4g_r^3k_r^3+g_r k_r^5\, . \label{ap3}
\end{eqnarray}\\

$p=7:$
\begin{eqnarray}
\fl f_{r+1}&=&f_r^7+6f_r^5g_r^2 +5f_r^4g_r^2k_r+10f_r^3g_r^4 +4f_r^3g_r^2k_r^2+12f_r^2 g_r^4k_r+4f_r g_r^6+3f_r^2 g_r^2k_r^3\nonumber \\
\fl&+&9f_r g_r^4k_r^2+3g_r^6k_r+2f_r g_r^2k_r^4+4g_r^4k_r^3+g_r^2k_r^5\, ,   \nonumber\\
\fl h_{r+1}&=&f_r^6g_r+f_r^5g_r k_r+5f_r^4 g_r^3 +f_r^4 g_r k_r^2+8f_r^3 g_r^3k_r+6f_r^2 g_r^5+f_r^3 g_r k_r^3+9f_r^2 g_r^3k_r^2 \nonumber \\
\fl &+&9f_r g_r^5 k_r + f_r^2g_r k_r^4+8f_r g_r^3k_r^3+g_r^7+6g_r^5k_r^2+f_r g_r k_r^5 +5g_r^3k_r^4+g_r k_r^6\, . \label{ap4}
\end{eqnarray}\\

$p=8:$
\begin{eqnarray}
\fl f_{r+1}&=&f_r^8+7f_r^6g_r^2 +6f_r^5g_r^2k_r+15f_r^4g_r^4 +5f_r^4g_r^2k_r^2+20f_r^3 g_r^4k_r+10f_r^2 g_r^6+4f_r^3 g_r^2k_r^3\nonumber \\
\fl&+&18f_r^2g_r^4k_r^2+12f_r g_r^6k_r+g_r^8+3f_r^2 g_r^2k_r^4+12f_r g_r^4k_r^3+6g_r^6 k_r^2+2f_r g_r^2k_r^5\nonumber \\
\fl&+&5 g_r^4k_r^4+ g_r^2k_r^6\, ,\nonumber \\
\fl h_{r+1}&=&f_r^7g_r+f_r^6g_r k_r+6f_r^5 g_r^3 +f_r^5 g_r k_r^2+10f_r^4 g_r^3k_r+16f_r^3 g_r^5+f_r^4 g_r k_r^3+12f_r^3 g_r^3k_r^2 \nonumber \\
\fl &+& 16f_r^2 g_r^5 k_r +f_r^3g_r k_r^4+12f_r^2 g_r^3 k_r^3+18f_r g_r^5k_r^2+4g_r^7k_r+f_r^2 g_r k_r^5 +10f_r g_r^3k_r^4\nonumber \\
\fl &+&10g_r^5 k_r^3 +f_r g_r k_r^6+6g_r^3 k_r^5+g_r k_r^7\, . \label{ap5}
\end{eqnarray}
As one can see from  equations~(\ref{ap1})-(\ref{ap5}) the number of terms and their coefficients increase rapidly  with  $p$.   Analyzing the model on MR  lattices with arbitrary $p$, we could formally write down recurrence equation  for $f$ in the following form
\begin{equation}\label{fp}
f_{r+1}=f_r^p+\sum_{i=1}^{p-1}\sum_{j=0}^{n_{i}}c_{ij}f^{\alpha_{ij}}g^{\beta_{ij}}k^{\gamma_{pij}}\, ,
\end{equation}
 where the upper limit   of the second sum,  the coefficients,  and the exponents,   all depend on $p$.
  \begin{figure}
\begin{center}
\includegraphics[scale=0.65]{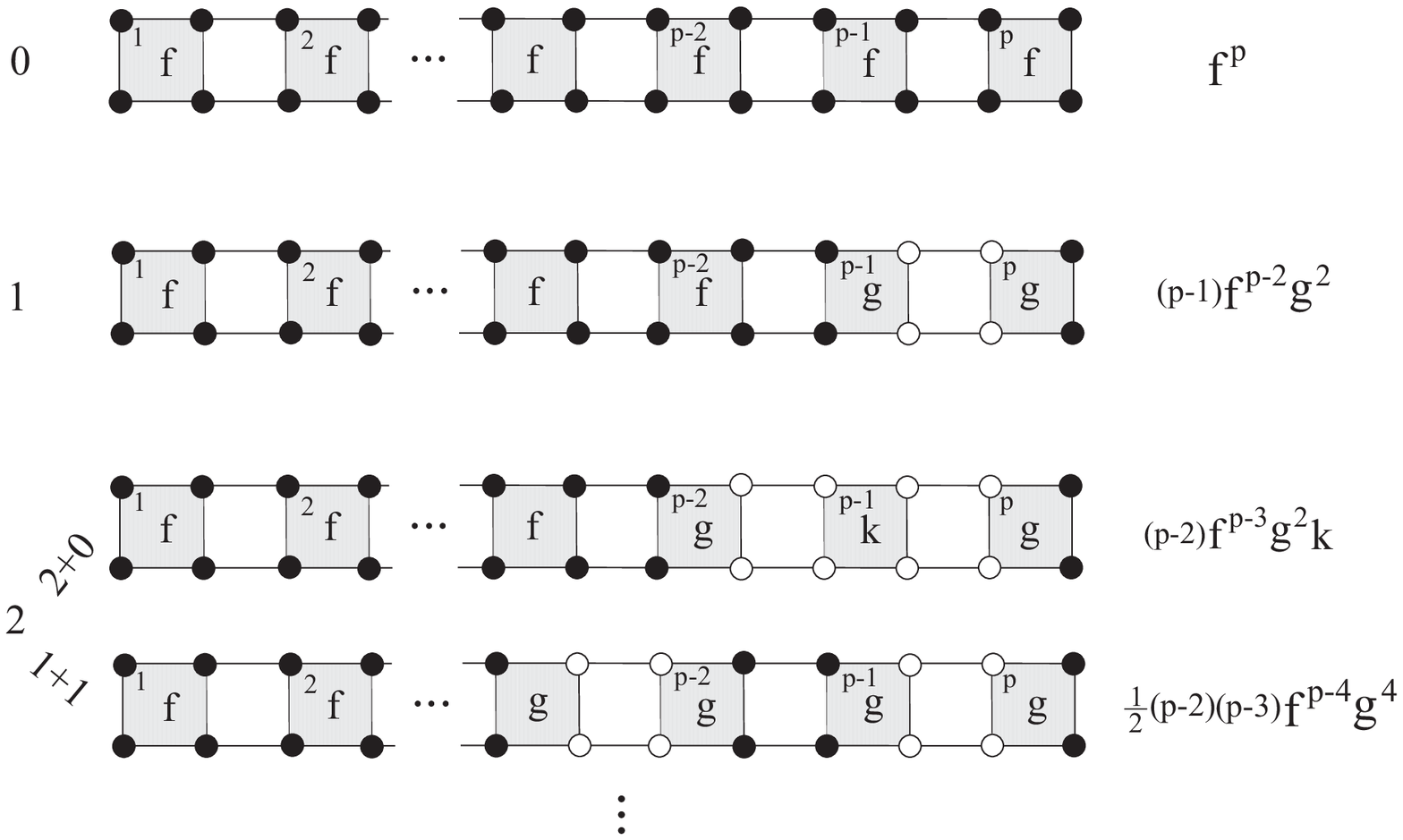}
\end{center}
\caption{ First few terms and their coefficients in recurrence equation~(\ref{fp}) for the total number of dimers $f_r$ on  MR lattice of large, arbitrary $p$.}
 \label{fig12}
\end{figure}
  The first term is obvious, and represents the only one possible way to obtain $f$ configuration  on $G_{r+1}$ composed from  $f$ configurations on each $G_r$, and this configuration is shown in the first row of figure~\ref{fig12}. The double  sum in equation (\ref{fp}) is  nontrivial. The first sum  represents terms that correspond to  situations in which one, two, three and so on 'empty' rectangles, with all four corner monomers white, can be chosen out of $(p-1)$ such rectangles, and, the second sum distinguishes whether these rectangles are   consecutive or not.  There are $(p-1)$ ways to choose one 'empty'  rectangle, so that coefficient $c_{10}$ of that term in the sum is $(p-1)$, (second row in figure~\ref{fig12}). Then, two consecutive rectangles can be chosen in $(p-2)$ ways, which gives $c_{20}=p-2$. Two non-consecutive rectangles can be chosen in $c_{21}=\frac{1}{2}(p-2)(p-3)$ ways (third row in figure~\ref{fig12}).    Three rectangles can  all be  consecutive, one separated and two consecutive and all three non-consecutive. Thus,  the upper limit of the second sum $n_{i}$ is  the number of ways in which a positive integer can be represented as the sum, which in combinatorics is known as partition.  There is no  simple formulae  for the partition function of a positive integer,  but it is known that  asymptotically it grows exponentially with the square root of its argument.  Therefore, the number of terms   increases very fast, and even if we succeeded to enumerate and sort the terms in recurrence equations, they would be    too  cumbersome to  analyze. At this point, we would like to notice that other methods might be more efficient  in handling the problem on the whole family of fractals, especially  in the large $p$ limit.

\end{document}